\documentclass[aps,prd,twocolumn,superscriptaddress,nofootinbib,preprintnumbers]{revtex4-1}
\usepackage{amsmath}
\usepackage{graphicx}
\usepackage{subfigure}
\usepackage{amssymb}
\usepackage{xcolor}
\usepackage{multirow}
\usepackage{cancel}
\usepackage{color}
\usepackage{ulem}
\usepackage{listings}
\usepackage{xcolor}
\usepackage{float}
\usepackage{slashed}

\usepackage{wrapfig}
\usepackage{mathtools}

\usepackage{tikz}
\usetikzlibrary{arrows,shapes}
\usetikzlibrary{trees}
\usetikzlibrary{matrix,arrows} 				
\usetikzlibrary{positioning}				
\usetikzlibrary{calc,through}				
\usetikzlibrary{decorations.pathreplacing}  
\usepackage{pgffor}							

\usetikzlibrary{decorations.pathmorphing}	
\usetikzlibrary{decorations.markings}
\tikzset{
    vector/.style={decorate, decoration={snake}, draw},
	provector/.style={decorate, decoration={snake,amplitude=2.5pt}, draw},
	antivector/.style={decorate, decoration={snake,amplitude=-2.5pt}, draw},
    fermion/.style={draw=black, postaction={decorate},
        decoration={markings,mark=at position .55 with {\arrow[draw=black]{>}}}},
    fermionbar/.style={draw=black, postaction={decorate},
        decoration={markings,mark=at position .55 with {\arrow[draw=black]{<}}}},
    fermionnoarrow/.style={draw=black},
    gluon/.style={decorate, draw=black,
        decoration={coil,amplitude=4pt, segment length=5pt}},
    scalar/.style={dashed,draw=black, postaction={decorate},
        decoration={markings,mark=at position .55 with {\arrow[draw=black]{>}}}},
    scalarbar/.style={dashed,draw=black, postaction={decorate},
        decoration={markings,mark=at position .55 with {\arrow[draw=black]{<}}}},
    scalarnoarrow/.style={dashed,draw=black},
    electron/.style={draw=black, postaction={decorate},
        decoration={markings,mark=at position .55 with {\arrow[draw=black]{>}}}},
	bigvector/.style={decorate, decoration={snake,amplitude=4pt}, draw},
}

\tikzstyle{block} = [draw, rectangle, 
    minimum height=3em, minimum width=6em]

\usepackage[colorlinks,citecolor=blue]{hyperref}
\usepackage{amsmath}
\usepackage{wrapfig}

\usepackage[T1]{fontenc} 
\usepackage[utf8]{inputenc} 
\usepackage{times}

\newcommand{\be}{\begin{equation}}
\newcommand{\ee}{\end{equation}}
\newcommand{\beq}{\begin{equation}}
\newcommand{\eeq}{\end{equation}}
\newcommand{\bea}{\begin{eqnarray}}
\newcommand{\eea}{\end{eqnarray}}
\newcommand{\besp}{\begin{equation}\begin{split}}
\newcommand{\eesp}{\end{split}\end{equation}}

\newcommand{\nn}{\nonumber}

\newcommand{\Eq}[1]{Eq.~(\ref{#1})}

\newcommand{\Dfbd}{\mathord{\buildrel{\lower3pt\hbox{$\scriptscriptstyle\leftrightarrow$}}\over {D}_{\mu}}}
\newcommand{\ave}[1]{\left\langle #1\right\rangle}

\newcommand{\figuretag}[1]{%
\addtocounter{figure}{-1}%
\renewcommand{\thefigure}{#1}%
}

\def\mL{\mathcal{L}}

\def\mN{\mathcal{N}}
\def\mO{\mathcal{O}}

\def\Z{\mathbb{Z}}

\def\0{\textbf{0}}
\def\1{\textbf{1}}
\def\2{\textbf{2}}
\def\3{\textbf{3}}
\def\4{\textbf{4}}
\def\5{\textbf{5}}
\def\6{\textbf{6}}
\def\7{\textbf{7}}
\def\8{\textbf{8}}
\def\9{\textbf{9}}

\def\d{\text{d}}
\def\hc{\text{h.c.}}
\usepackage{fontawesome5}

\begin{document}

\title{Asteroid-mass soliton as the dark matter-baryon coincidence solution}

\author{Shinya Kanemura}
\email{kanemu@het.phys.sci.osaka-u.ac.jp}
\affiliation{Department of Physics, The University of Osaka, Toyonaka, Osaka 560-0043, Japan}

\author{Shao-Ping Li}
\email{lisp@het.phys.sci.osaka-u.ac.jp}
\affiliation{Department of Physics, The University of Osaka, Toyonaka, Osaka 560-0043, Japan}
	
\author{Ke-Pan Xie}
\email{kpxie@buaa.edu.cn}
\affiliation{School of Physics, Beihang University, Beijing 100191, P. R. China}

\preprint{OU-HET-1272} 
	
\begin{abstract}
Nontopological solitons formed during first-order phase transitions can serve as macroscopic dark matter candidates, with their stability ensured by a charge asymmetry traditionally assumed to originate from baryogenesis. Following this generic pattern, we demonstrate that solitogenesis after baryogenesis makes the solitons a coincident dark matter candidate, providing new explanations for the coincidence problem between baryon and dark matter energy densities. We derive a novel and robust conclusion: asteroid-mass coincident soliton dark matter is always accompanied by detectable gravitational waves observable by LISA, $\mu$Ares, and Theia, providing a new candidate beyond primordial black holes in this mass window. Additionally, we propose a simple neutrino-ball scenario that addresses baryon asymmetry, dark matter, and neutrino masses, featuring new particles below the electroweak scale and correlated observable signals, including lensing, gravitational waves, and soliton evaporation or collisions.
\end{abstract}
\maketitle

\textbf{Introduction}. The Standard Model (SM) in elementary particle physics has demonstrated itself as a minimal and self-consistent theory in explaining and predicting numerous observational facts. Being tremendously successful, nevertheless, the SM leaves several open questions, including the dynamical explanation of the baryon asymmetry in the Universe (BAU), the candidates of dark matter (DM), and the origin of neutrino masses. Additionally, the DM energy density ($\Omega_{\rm dm}=0.265$) and the baryon energy density ($\Omega_{\rm b}=0.0493$) present a coincidence problem $\Omega_{\rm dm}/\Omega_{\rm b}\approx 5.4$~\cite{Planck:2018vyg}, suggesting  an underlying common origin~\cite{Velten:2014nra}.

Nontopological solitons~\cite{Zhou:2024mea,Nugaev:2019vru,Lee:1991ax,Friedberg:1976eg,Friedberg:1977xf,Koeppel:1985tt,Lee:1986tr,Rosen:1968mfz,Friedberg:1976me,Coleman:1985ki} are macroscopic DM candidates~\cite{Jacobs:2014yca}, which typically exist as compact balls confining numerous fermions (Fermi-balls~\cite{Hong:2020est,Marfatia:2021twj,Gross:2021qgx,Kawana:2021tde,Kawana:2022lba,Chakrabarty:2024pvf,Xie:2024mxr,Lu:2024xnb}, also called quark nuggets when the constituents are quark-like~\cite{Witten:1984rs,Bai:2018vik,Bai:2018dxf,Bai:2024muo,Liang:2016tqc,Ge:2017idw,Ge:2019voa,Zhitnitsky:2021iwg}) or scalar bosons (Q-balls~\cite{Krylov:2013qe,Huang:2017kzu,Bai:2022kxq,Carenza:2024tmi,Jiang:2024zrb}). Generally, nontopological soliton DM necessitates a charge asymmetry, as its stability is guaranteed by the nonzero conserved N\"{o}ther charge carried by the constituent particles. This charge asymmetry has been widely considered to be of the same order as the baryon asymmetry (see e.g., the aforementioned references), which implies an underlying connection between soliton DM and the BAU. This assumption may open new avenues to explain the coincidence problem by macroscopic objects rather than the conventional particle DM paradigm~\cite{Barr:1990ca,Kaplan:1991ah,Barr:1991qn,Kuzmin:1996he,Kitano:2004sv,Farrar:2005zd,Hooper:2004dc,Kaplan:2009ag}. 

Following this generic pattern, we demonstrate in this Letter that when the BAU and the soliton charge asymmetry share a common origin,  solitogenesis from a later first-order phase transition (FOPT) can explain the coincidence problem. Such \textit{coincident soliton DM} has  a robust feature that deserves highlight: if it resides in the  asteroid mass scale, $10^{12}~\text{g}-10^{22}$~g, strong gravitational waves (GWs) generated during the FOPT will always reach the detection regions of LISA~\cite{LISA:2017pwj}, $\mu$Ares~\cite{Sesana:2019vho}, or Theia~\cite{Theia:2017xtk}. This feature is general and independent of how baryogenesis provides the initial charge asymmetry for solitons. 
In addition, coincident soliton DM provides a new alternative to the primordial black holes (PBHs)~\cite{Hawking:1971ei,Carr:1974nx,Chapline:1975ojl,Carr:2020xqk,Green:2020jor,Carr:2020gox} in the mass range $10^{17}~\text{g}-10^{22}$~g~\cite{Macho:2000nvd,Wilkinson:2001vv,EROS-2:2006ryy,Griest:2013esa,Griest:2013aaa,Oguri:2017ock,Zumalacarregui:2017qqd,Niikura:2019kqi,Smyth:2019whb}, where several  experimental concepts and targets, such as femtolensing/picolensing~\cite{Nemiroff:1995ak,Kolb:1995bu,Marani:1998sh,Katz:2018zrn,Jung:2019fcs,Gawade:2023gmt,Fedderke:2024wpy} and time information~\cite{Kaplan:2024dsn} of gamma-ray bursts, as well as microlensing of X-ray pulsars~\cite{Bai:2018bej}, are under active development. Moreover, the complementary signals from GWs and lensing effects in the mass range $10^{17}~\text{g}-10^{22}$~g can also be used to distinguish  coincident soliton DM from PBHs.

As an application, we provide a simple and realistic neutrino-ball scenario based on the Dirac seesaw~\cite{Roncadelli:1983ty,CentellesChulia:2016rms} to realize coincident Fermi-ball DM, which can  simultaneously address the BAU, DM, and neutrino masses with all the beyond-SM particles  below the electroweak scale, allowing for direct detection at colliders.

\begin{figure}[b]
	\centering
	\includegraphics[scale=0.27]{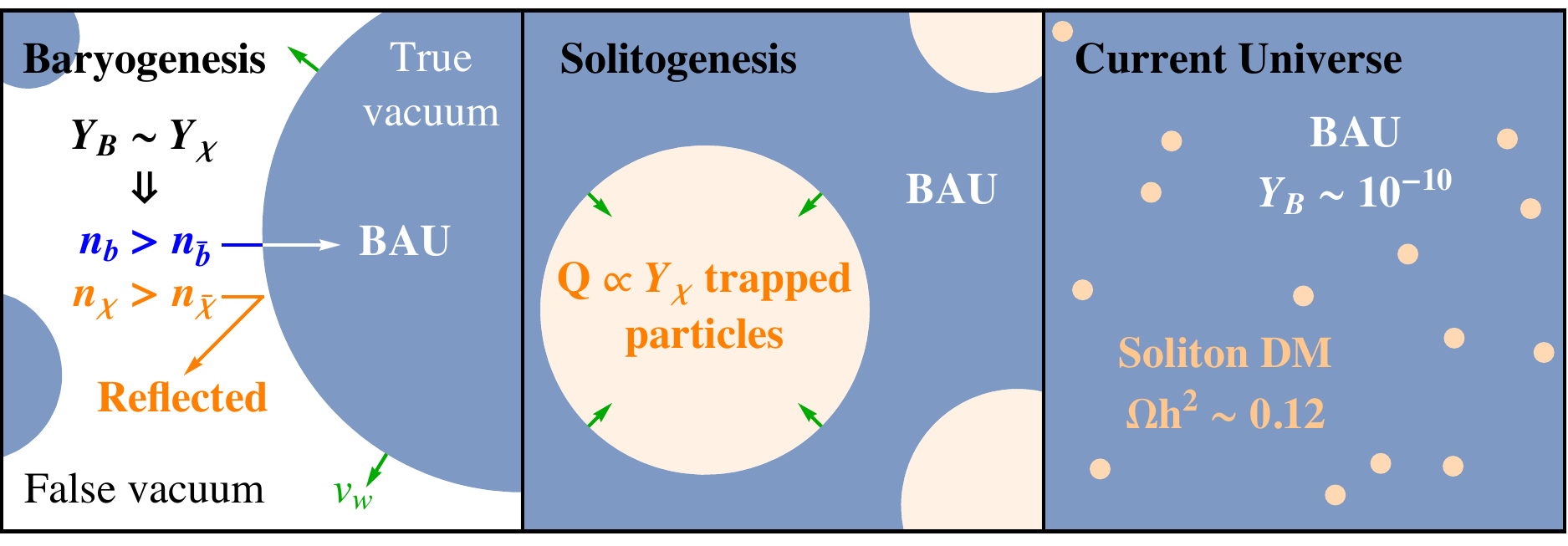} 
	\caption{Sketch of the paradigm. Both the baryon asymmetry ($Y_B$) and the hidden sector charge asymmetry ($Y_\chi$) are generated during the baryogenesis process, satisfying  $Y_\chi/Y_B\sim\mO(0.1-1)$. The SM particles penetrate into the true vacuum, yielding the observed BAU. The asymmetric dark particles inherited from baryogenesis are reflected by the bubble wall due to their large mass gap, consequently being trapped in the false vacuum to form soliton DM.}\label{fig:sketch}
\end{figure}

\textbf{Coincident Fermi-ball DM}. For definiteness, let us consider the interaction between the SM and hidden-sector fermions $\chi$. It could be the lepton portal $\bar\ell_L\tilde H\chi_R$, with $\ell_L$ and $H$ the left-handed lepton and Higgs doublets, respectively, or the quark portal $\bar q_L S\chi_R$, with $q_L$ the left-handed quark doublet and $S$ the leptoquark in grand unified theory or squark in supersymmetric theory.

\begin{figure*}[t]
	\centering
	\includegraphics[scale=0.5]{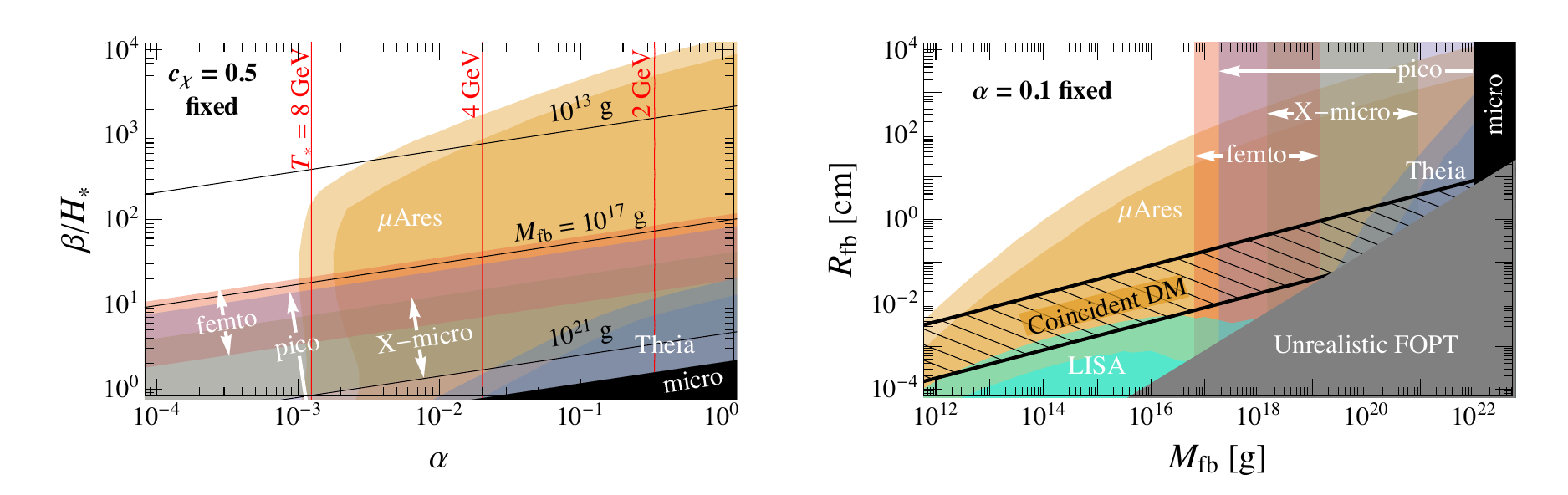} 
	\caption{The parameter space of coincident Fermi-ball DM. Left: $T_*$ and $M_{\rm fb}$ as functions of $(\alpha,\beta/H_*)$, with $c_\chi=0.5$ fixed. Right: $c_\chi$ as a function of $(M_{\rm fb},R_{\rm fb})$, with $\alpha=0.1$ fixed. The colored shaded regions indicate the constraints and projected detection regimes from   lensing and GW experiments. See the text for details.}\label{fig:landscape}
\end{figure*}

The generic pattern is  sketched in Fig.~\ref{fig:sketch}. The $\chi$-asymmetry can be written as $Y_\chi\equiv (n_\chi-n_{\bar\chi})/s=c_\chi Y_B$, where $Y_B\equiv   (n_b-n_{\bar b})/s$ denotes the baryon asymmetry normalized to the entropy density $s=g_{\rm eff} 2\pi^2T^3 /45$   with $g_{\rm eff}$ the effective degrees of freedom, and  $c_\chi$ is an $\mO(1)$ coefficient depending on the specific realization of baryogenesis.

After baryogenesis, $\chi$ particles experience FOPT from some background scalar field $\phi$ and develop a mass gap via an interaction $y_\phi\phi\bar\chi\chi$, with $y_\phi$ the Yukawa coupling. If this mass gap is much larger than the FOPT temperature $T_*$, the $\chi$-asymmetry can result in a number of $Q\approx Y_\chi s_*(4\pi R_*^3/3)$ trapped in the false vacuum remnants,   where $s_*$ is the entropy density at formation epoch, and $R_*\approx v_w/\beta$~\cite{Lu:2022paj} is the size of the remnants with $v_w$ the bubble wall velocity and $\beta^{-1}$ the FOPT duration. The remnants eventually shrink to a size much smaller than $R_*$ and form individual Fermi-balls.

After formation, Fermi-balls track the Universe's temperature and cool down~\cite{Bai:2018dxf,Hong:2020est,Kawana:2022lba}. As the profile depends on the temperature weakly~\cite{Kawana:2021tde}, one can approximate  the Fermi-ball mass and radius respectively as $M_{\rm fb}=Q(12\pi^2V_0)^{1/4}$ and $R_{\rm fb}=[3M_{\rm fb}/(16\pi V_0)]^{1/3}$ by using  the present-day  profile, where $V_0$ is the free energy difference between the true  and false vacua. The number density of Fermi-balls is the same as that of the remnants at the moment of formation, $n_{\rm fb}^*\sim R_*^{-3}$, and then evolves by entropy conservation under the cosmic expansion.

Using these features, one can obtain the energy ratio of soliton DM to baryons as
\be\label{eq:cos-con}
\frac{\Omega_{\rm dm}}{\Omega_{\rm b}}\approx c_\chi\left(\frac{\mu}{m_p}\right),
\ee
where $m_p\approx0.938$ GeV is the proton mass, $\mu\equiv\text{d}M_{\rm fb}/\text{d}Q$ is the effective mass for a constituent $\chi$ inside the soliton. The above relation shows that  an $\mO(1)$ $c_\chi$ and a GeV-scale $\mu$ explain why the energy densities of DM and baryons  have the similar order, even though they have very different particle physics backgrounds. This differs from the  asymmetric particle DM paradigm for the coincidence problem, since $\mu$ is not the physical DM mass but is induced nontrivially by the FOPT trapping the asymmetric $\chi$ particles. Here, the observable DM mass $M_{\rm fb}$ provides a connection between the coincidence problem and the macroscopic mass window directly associated with the lensing effects.

For generic FOPT, $V_0\approx V_0^*\approx(\pi^2g_*T_*^4/30)\alpha$ holds to a good approximation, where $\alpha$ denotes the ratio of FOPT latent heat to the radiation energy density. For Fermi-ball contributing to the whole DM relic density, the FOPT temperature reads $T_*\approx (0.1/\alpha)^{1/4}c_\chi^{-1}~{\rm GeV}$, suggesting a GeV-scale FOPT. Substituting $T_*$ into $M_{\rm fb}$, we find the Fermi-ball DM mass links the BAU and the FOPT dynamics via
\be
M_{\rm fb}\approx10^{17}~{\rm g}\left(\frac{v_w}{0.6}\right)^3\left(\frac{100}{\beta/H_*}\right)^3\left(\frac{\alpha}{0.1}\right)^{3/4}c_\chi^3,
\ee
 where we set $g_{\rm eff}=80$ as a benchmark. It suggests a strong correlation between the Fermi-ball mass and the GW signals, as the latter is determined by $v_w$, $\alpha$, and $\beta/H^*$. For typical FOPT with $\alpha=\mathcal{O}(0.1)$ and $\beta/H^*=\mathcal{O}(100)$, the GW is sound-wave dominated~\cite{Caprini:2019egz}. The resulting GW peak frequency and amplitude both depend on the Fermi-ball mass,
\bea
f_{\rm sw}&\approx&3.8\times10^{-5}~{\rm Hz}\left(\frac{10^{17}~{\rm g}}{M_{\rm fb}}\right)^{1/3},
\\
\Omega_{\rm sw}&\approx&10^{-11}\left(\frac{M_{\rm fb}}{10^{17}~{\rm g}}\right)^{2/3}\left(\frac{\alpha}{0.1}\right)\left(\frac{\mu}{m_p}\right)^2\left(\frac{\Omega_{\rm b}}{\Omega_{\rm dm}}\right)^2.\nn
\eea
Remarkably, the robust prediction that asteroid-mass Fermi-ball DM is  associated with micro-Hertz GWs perfectly aligns with the detection targets of the proposed experiments $\mu$Ares and Theia, providing another theoretical motivation  to advance the development of these detectors.

Fig.~\ref{fig:landscape} illustrates the parameter space for coincident soliton DM alongside projected experimental sensitivities. The left panel displays contours of $M_{\rm fb}$ and $T_*$ as functions of FOPT parameters $\alpha$ and $\beta/H_*$, with $c_\chi=0.5$ fixed. After the FOPT, the Universe is reheated to a temperature $T_{\rm rh}\approx(1+\alpha)^{1/4}T_*$. This might melt the solitons, if $T_{\rm rh}$ is comparable with the mass gap, such that $\chi$'s gain sufficient kinetic energy to penetrate the true vacuum. Therefore, $\alpha<1$ is required to have a moderate FOPT,  while $\beta/H_*<1$ is not considered as it is not realizable in general FOPTs. The right panel shows the $M_{\rm fb}$-$R_{\rm fb}$ plane fixing $\alpha=0.1$, where the meshed region represents the natural regime $5~{\rm GeV}\lesssim\mu\lesssim50$ GeV for the coincidence problem, corresponding to $c_\chi\simeq 0.1-1$. In both panels, $v_w=0.6$, and the orange, blue and cyan shaded regions indicate the projected GW sensitivities of future $\mu$Ares~\cite{Sesana:2019vho}, Theia~\cite{Theia:2017xtk}, and LISA~\cite{LISA:2017pwj} detectors, respectively, with a  signal-to-noise ratio at 10 (lighter) and 100 (darker). Existing constraints from  microlensing~\cite{Carr:2020xqk,Green:2020jor,Carr:2020gox} and projected detection regimes~\cite{Katz:2018zrn,Bai:2018bej,Jung:2019fcs,Gawade:2023gmt,Fedderke:2024wpy} are also shown for reference. Different from the PBH case, Fermi-ball does not suffer from Hawking radiation~\cite{Hawking:1974rv,Hawking1975}, and hence the coincidence solution allows for mass as light as $\sim 10^{12}$ g. However, the combined detection of GWs and lensing restricts in the asteroid-mass window, which provides an important distinguished signature between Fermi-balls and PBHs.

\textbf{Lifetime}. The solitons are stable against evaporating into free $\chi$ fermions if $\mu$ is smaller than the free particle mass $m_\chi=y_\phi \langle \phi\rangle$ in the true vacuum, and are stable against splitting into two smaller solitons if $\d^2M_{\rm fb}/\d Q^2\leqslant0$. However, due to the portal coupling required by the generation of the BAU and $\chi$-asymmetry, the constituent $\chi$ fermions near the Fermi-ball surface may decay to SM light particles, leading to soliton  evaporation~\cite{Cohen:1986ct}. Nevertheless, the soliton lifetime due to particle decay through surface evaporation can be    longer than the age of the Universe, since the constituents deep inside the soliton are massless. To see this, one can assume that only the particles with an effective mass $\mu$ within the bubble width $\sim m_\phi^{-1}$ can decay. The lifetime of a soliton estimated via $\tau_{\rm fb}\sim -Q/(\d Q/\d t)$ yields 
\be\label{eq:lifetime}
\frac{\tau_{\rm fb}}{\tau_\chi}\approx10^{13}\left(\frac{m_\phi}{\rm GeV}\right)\left(\frac{\Gamma_\chi}{\Gamma_\chi^\mu}\right)\left(\frac{M_{\rm fb}}{10^{17}~{\rm g}}\right)^{1/3}c_\chi^{4/3},
\ee
where $\Gamma_\chi^\mu$ ($\Gamma_\chi$) is the decay rate evaluated with  mass $\mu$ ($m_\chi$). Given that $\mu<m_\chi$, $\Gamma_\chi^\mu$ is   smaller than $\Gamma_\chi$, especially when $\mu$ is smaller than the true vacuum masses of the decay products such that the decay channels are of multi-body type. Therefore, we see that the lifetime of a Fermi-ball is enhanced significantly compared with that of a free $\chi$ fermion. Depending on the particle physics models, the Fermi-ball lifetime longer than the age of the Universe $\tau_{\rm fb}\gtrsim 10^{17}$~s can be easily realized by a small $\Gamma_\chi^\mu$.

It is worth mentioning that constraints from soliton DM evaporation are weaker than from decaying particle DM~\cite{Cirelli:2012ut,Essig:2013goa,Blanco:2018esa}. The injected particle flux and energy spectrum from decaying DM is well determined  by the particle DM mass and the (local) relic density, however, soliton DM decay  is  only through surface evaporation, where the   energy spectrum of products  only results from a small fractional mass of a single Fermi-ball. Determining the spectrum of the injected particles from soliton evaporation or collision of solitons~\cite{Bai:2018dxf} would provide a distinctive feature from decaying DM and   a complementary test for  Fermi-ball DM. While the precise calculation of Fermi-ball evaporation is still challenging, it can open a new research direction  and deserves careful consideration in future studies. 

\textbf{Neutrino-balls}. As a concrete application of the proposed pattern, let us identify the fermions $\chi$ as the sterile neutrinos. Majorana neutrinos  are a leading candidate for explaining the SM neutrino masses, but  it is challenging to make them Fermi-ball constituents via FOPT due to annihilation problems. Confining $\chi$ into a soliton requires interactions with a background field  to create a space-dependent fermion mass, trapping them via the mass gap. For Majorana $\chi$, the corresponding interaction $\phi\chi\chi$ inevitably leads to annihilation of $\chi$ pairs to scalar quanta via $\chi \chi\to\phi$, $\chi \chi\to\phi\phi$. This would eventually lead to the $\chi$-disappearance even if there is an initial helicity asymmetry for Majorana $\chi$, which is a consequence previously overlooked~\cite{Holdom:1987ep,Dolgov:1990sr}. For Dirac $\chi$ with interactions $\phi\bar\chi\chi$, in contrast, an initial charge asymmetry can be protected from pair annihilation, leading to net Dirac $\chi$ particles surviving to be soliton constituents.

While a substantial number of scenarios for Dirac neutrinos have been developed, we here concentrate on the minimal framework that allows  stronger correlations among various observational consequences. To this end, we 
consider the Dirac type-I seesaw mechanism~\cite{Roncadelli:1983ty,CentellesChulia:2016rms} with the Lagrangian
\be\label{Dirac_seesaw_sim}
\mL= -y_h\bar\ell_L\tilde H\chi_R-y_\eta\bar \chi_L\eta\nu_R-y_\phi \phi\bar \chi_L\chi_R+\hc
\ee
introduced to the SM, where $\nu_R$ is the Dirac right-handed counterpart of the SM active neutrinos. The sterile neutrinos $\chi$  are  assumed to have three  generations. Two real singlet scalars $\eta$ and $\phi$ help the asymmetry redistribution among three $\chi$ flavors, serve to trigger the FOPT, and explain the mass origins for the neutrinos.  We use  $y_{i}$ to denote the Yukawa couplings associated with different scalars $i=h$, $\eta$, and $\phi$. After the scalars develop their vacuum expectation values (VEVs), i.e., $\ave{h}\equiv v\approx 246$~GeV, $\ave{\phi}\equiv w$, and $\ave{\eta}\equiv u$, the SM neutrino masses read 
\begin{align}\label{eq:nu-mass}
m_\nu\approx17~{\rm meV} \left(\frac{y_h}{10^{-6}}\right)\left(\frac{y_\eta}{10^{-4}}\right)\left(\frac{u}{10^{-3}m_\chi}\right).
\end{align}
Note a hierarchy $u\ll m_\chi$ is implied in Dirac seesaw: to forbid the direct mass term $\bar\ell_L\tilde H\nu_R$, a $\Z_2$ symmetry is assigned, under which $\nu_R$ and $\eta$ are odd; however, there should be a soft $\Z_2$-breaking term in the scalar potential, such that $\eta$ obtains a small VEV to realize \Eq{eq:nu-mass}. This naturally leads to the formation of $\eta$-based domain wall in the early Universe~\cite{Barman:2022yos}, which is also essential for our mechanism, to be detailed below.

The first event for realization of coincident neutrino-ball DM is leptogenesis that happens before the electroweak phase transition and $\phi$-FOPT, where $\ave{h}=\ave{\phi}=\ave{\eta}=0$. For minimality and simplicity, we build on the idea of the shared and conserved $B-L$ between the SM and hidden sector particles~\cite{Kaplan:1991ah,Barr:1991qn,Kuzmin:1996he,Akhmedov:1998qx,Dick:1999je,Davoudiasl:2010am,Kitano:2004sv,Farrar:2005zd,Elor:2018twp,Kanemura:2024dqv}, where  the dark asymmetry and BAU are both generated via leptogenesis.  Leptogenesis then proceeds through the out-of-equilibrium decay $H\to\bar\ell_L\chi_R$ via the first term in \Eq{Dirac_seesaw_sim}. As the same portal also causes $\chi$ decay  and hence neutrino-ball   evaporation,  we assume $\chi_1$ has weaker interactions such that   the BAU is dominantly generated via    Higgs decay to  $\chi_{2,3}$. Finite-temperature corrections to the SM leptons will induce CP asymmetries equally distributed in the SM lepton and $\chi$ sectors~\cite{Beneke:2010dz,Garbrecht:2012pq,Kanemura:2024dqv,Kanemura:2024fbw}, where the CP-violating source for generating the $\chi$-asymmetry exhibits a resonant enhancement. Following  Ref.~\cite{Kanemura:2024dqv}, the asymmetry for $\chi_2+\chi_3$ yields 
 \begin{align}\label{eq:Ynu}
 Y_{ \chi_{2}+\chi_3} \approx4.3\times 10^{-8}\left(\frac{\bar y_{h}}{10^{-6}}\right)^4\langle f_{\chi}\rangle_{\alpha\beta}\,,
 \end{align}
where $\bar y_h\equiv\sum_{\alpha,\beta=2,3}  [\text{Im}(y_{h, e\alpha}\,y^*_{h,\mu\alpha}\,y_{h,\mu\beta}\,y^*_{h,e\beta})]^{1/4}$ with the indices $\mu$ and $e$ corresponding to the maximal resonant enhancement in the muon-electron direction~\cite{Kanemura:2024dqv}.  The average $\langle f_{\chi}\rangle_{\alpha\beta}$ is defined by the nonthermal $\chi_{2,3}$ distribution functions under interaction over temperature and momenta, 
\begin{align}
	\langle f_{\chi}\rangle_{\alpha\beta}&\equiv \int_0^{z_{\rm sph}}\d z \int_0^\infty \d x\int_{y_{\rm low}}^\infty \d y\int_{w_{\rm low}}^\infty \d w\frac{\theta_z^4 I_{\alpha\beta}}{x}\,,
\end{align}
where $z\equiv m_h/T$, and $\theta_z\equiv [z^2\theta(z-z_c)+0.3]^{1/2}$ results from  the thermal mass difference between the Higgs and leptons, with $z_c\approx 0.78$ corresponding to the electroweak crossover~\cite{DOnofrio:2014rug}.  $z_{\rm sph}\approx 0.95$ denotes the end of sphaleron conversion, while $y_{\rm low}\equiv \theta_z^2/(4x)+x$ and $w_{\rm low}\equiv \theta_z^2/(4x)$ result from kinematic thresholds.   The statistics function is $I_{\alpha \beta}\equiv  [f_{\chi_\beta}(w)-f_{\chi_\beta}^{\rm eq}(w)][f_H^{\rm eq}(y)-f_{\chi_\alpha}(y-x)][f_H^{\rm eq}(w+x)-f_{\ell}(x)]$ with $f^{\rm eq}$   the equilibrium distribution.  

We estimate the evolution of $f_\chi$ via Higgs decay and inverse decay for simplicity, though including gauge interactions can lead to an $\mathcal{O}(1)$ enhancement~\cite{Anisimov:2010gy,Besak:2012qm}. The generated $Y_{\chi_2+\chi_3}$ causes a SM lepton asymmetry $Y_{L_{\rm SM}}=-Y_{\chi_2+\chi_3}$, which is then partly converted into the baryon asymmetry via the electroweak sphaleron processes $B=c_{\rm sph} (B-L_{\rm SM})$ with $c_{\rm sph}$ the conversion efficiency~\cite{Harvey:1990qw}. Assuming a maximal CP violation from the Yukawa couplings, we found that  $y_{h,e\alpha},\, y_{h,\mu\alpha}\simeq10^{-6}$ can explain the observed   BAU $Y_B\approx0.9\times10^{-10}$~\cite{Planck:2018vyg}. After the sphaleron processes decouple at 132 GeV~\cite{DOnofrio:2014rug}, the baryon asymmetry is frozen, and a net lepton asymmetry  $Y_L=Y_B$ will be left in the thermal bath due to $B-L$ conservation.  This net $Y_L$ will eventually be redistributed among  $\ell,  \nu, \chi$ flavors,   particularly with $\chi_1$. Such redistribution processes proceed via    the thermalized Yukawa interactions from the second and third terms of \Eq{Dirac_seesaw_sim} after the sphaleron decoupling.  Resolving chemical potential equilibrium equations with three thermalized $\nu_R$ and $\chi$ flavors yields $Y_ {\chi_1}=8Y_B/63$. This explains the coincidence problem with  $c_\chi=8/63\approx0.127$, and provides the initial asymmetry for $\chi_1$ to form neutrino-balls during the FOPT.

 The second event is solitogenesis that follows   leptogenesis. The solitogenesis process forms Fermi-balls via trapping particles in the false vacuum remnants of cosmic first-order phase transitions (FOPTs), as proposed by Witten~\cite{Witten:1984rs} and then extensively studied~\cite{Hong:2020est,Marfatia:2021twj,Gross:2021qgx,Kawana:2021tde,Kawana:2022lba,Chakrabarty:2024pvf,Bai:2018vik,Bai:2018dxf,Bai:2024muo,Krylov:2013qe,Huang:2017kzu,Bai:2022kxq,Carenza:2024tmi,Jiang:2024zrb}. 
 In the simplest case that the scalar potential is even under both $\phi\to-\phi$ and $\eta\to-\eta$, the $\phi$-$\eta$ system can trigger a FOPT via the well-known two-step process, i.e.
\be\label{fopt}
(\phi=0,\eta\simeq0)\to(\phi=0,\eta\neq0)\to(\phi\neq0,\eta\simeq0)\,,
\ee
where the second step is a FOPT. Since the SM neutrino masses in \Eq{eq:nu-mass} requires $u\ll w_*$ without too feeble Yukawa couplings, the $\eta$ field favors an approximate $Z_2$ symmetry. In this case, we found it more effective to trigger the FOPT via the $\eta$-direction domain wall catalysis~\cite{Blasi:2022woz,Blasi:2023rqi} than the traditional homogeneous nucleation. In this scenario, the $\eta$-based domain walls form in the first step, which serve as local impurities for the bubble nucleation in the second step. We check this pattern numerically in the parameter space where the thin-wall approximation is valid. The results confirm that the FOPT can indeed occur with a GeV-scale $T_*$, and provide a mass gap $y_\phi\ave{\phi}\gtrsim10\,T_*$ to trap $\chi$ fermions and form neutrino-balls with a mass $M_{\rm nb}\simeq10^{20}$ g, which is in the asteroid-mass window. The calculation details are presented in the Appendix.

The  coefficient $c_\chi=0.127$ results from three thermalized right-handed Dirac neutrinos and $\chi$. This can be readily realized by the Yukawa matrix $y_\eta$ having elements at the similar order $y_\eta\simeq 10^{-4}$ and by the Yukawa matrix $y_h$ having hierarchical elements $y_{h, i1}\ll y_{h, i2}, y_{h, i3}\simeq 10^{-6}$, the latter of which is dictated by   successful leptogenesis via Higgs decay to $\chi_{2,3}$ and meanwhile by stable neutrino-balls made up of $\chi_1$. Such a flavor pattern suggests that the lightest neutrino in the SM should have a mass much smaller than the two heavier ones.  Moreover, the mass ratio of these SM neutrinos are correlated with the neutrino-ball lifetime. Due to the lepton portal coupling, the SM light neutrino flux in the Universe space can cause the volume evaporation via e.g. $\chi\nu\to\ell^+\ell^-$, however we have checked that this effect is negligible due to the relatively low SM neutrino number density. For $\mu<m_{\phi,\eta}<50$~GeV induced from successful neutrino-ball formation, the life time of neutrino-ball is dominated by surface evaporation via three-body decay $\chi_1\to \ell^+\ell^-\nu$, $3\nu$, with an effective mass $\mu$ for $\chi_1$. Based on Eq.~\eqref{eq:lifetime}, we found that a lifetime of $\tau_{\rm fb}=10^{18}$~s corresponds to a mass ratio of the lightest to the heavier neutrinos at $10^{-6}$. The longer the lifetime of the neutrino-balls, the smaller the mass ratio will be, a prediction that can be tested in upcoming neutrino experiments such as KATRIN~\cite{KATRIN:2021dfa} and cyclotron radiation emission from Project 8~\cite{Project8:2017nal}.

The three thermalized $\nu_R$ flavors also  contribute to   extra energy budget of the early Universe. Given $m_\phi, m_\eta\lesssim m_\chi$,  the decoupling temperatures of these $\nu_R$ can be estimated via decay $\chi\to \eta \nu_R$. For  $y_\eta\simeq 10^{-4}$ and $m_\chi\gtrsim 50$~GeV, the decoupling temperature lies between 2~GeV and 10 GeV. It  causes a modification of the effective neutrino number $\Delta N_{\rm eff}$ smaller than the current Planck bound $\Delta N_{\rm eff}<0.29$~\cite{Planck:2018vyg} but above $\Delta N_{\rm eff}=0.14$ that can be probed by Simons Observatory~\cite{SimonsObservatory:2018koc} and CMB-S4~\cite{CMB-S4:2016ple} experiments. 
 
\textbf{Conclusion}. We have demonstrated in this Letter that nontopological soliton DM  can explain the coincidence problem when the initial charge asymmetry is inherited from some baryogenesis process. We found a robust and general prediction from this coincident soliton DM pattern:  asteroid-mass soliton DM at  $10^{12}~\text{g}-10^{22}$~g is always accompanied with   strong GWs that will  reach the detection regions of  LISA, $\mu$Ares, and Theia, which is independent of how  baryogenesis provides the initial charge asymmetry for solitons. 
In addition, we have also provided a realistic neutrino-ball scenario, which is quantitatively realized for the first time in a minimal particle physics framework, where all the beyond-SM particles  are below the electroweak scale and hence make current and future collider detection attainable.

\textbf{Acknowledgements}. We would like to thank Simone Blasi, Shuailiang Ge, Wei Liu, Sida Lu, Yue-Lin Tsai, and Daneng Yang for the useful discussions. K.-P.~Xie also thanks the hospitality of the University of Osaka where part of this work was performed. S.~Kanemura and S.-P.~Li are supported by JSPS Grant-in-Aid for JSPS Research Fellows No. 24KF0060.  S.~Kanemura is also supported in part by the JSPS KAKENHI Grant No. 20H00160 and No. 23K17691. K.-P.~Xie is supported by the National Natural Science Foundation of China under Grant No. 12305108, and by the Fundamental Research Funds for the Central Universities.

\section*{Appendix: FOPT dynamics and neutrino-ball formation}
The FOPT in the Dirac type-I seesaw framework can be triggered by the two real scalar singlets, $\phi, \eta$, where  the effective scalar potential at temperature $T$ is given by 
\begin{align}\label{VT}
V_T(\phi,\eta,T)=\frac{\mu_\phi^2+c_\phi T^2}{2}\phi^2+\frac{\mu_\eta^2+c_\eta T^2}{2}\eta^2\nn\\
+\frac{\lambda_\eta}{4}\eta^4+\frac{\lambda_\phi}{4}\phi^4+\frac{\lambda_{\phi\eta}}{4}\phi^2\eta^2,
\end{align}
with  the   coefficients $c_\phi, c_\eta$ derived by thermal corrections from particles interacting with  the scalar fields
\be
c_\phi=\frac{\lambda_\phi}{4}+\frac{\lambda_{\phi\eta}}{24}+\frac{y_\phi^2}{2},\quad
c_\eta=\frac{\lambda_\eta}{4}+\frac{\lambda_{\phi\eta}}{24}.
\ee
The necessary condition for the two-step phase transition in \Eq{fopt} with the second step being a FOPT is~\cite{Bian:2019kmg}
\be\label{analytical_Tc}
\frac{c_\eta}{c_\phi}<\frac{\mu_\eta^2}{\mu_\phi^2}<\frac{\sqrt{\lambda_\eta}}{\sqrt{\lambda_\phi}}<\frac{\lambda_{\phi\eta}}{2\lambda_\phi}.
\ee
However, this is not the sufficient condition. To verify if the vacuum decay can really happen, we need to evaluate the tunneling rate and demonstrate that bubbles can nucleate and fulfill the whole Universe. For the parameter space under consideration, the domain-wall-catalyzed phase transition~\cite{Blasi:2022woz} is the dominant pathway to realize the FOPT. 

\begin{figure}[b]
\centering
\includegraphics[scale=0.5]{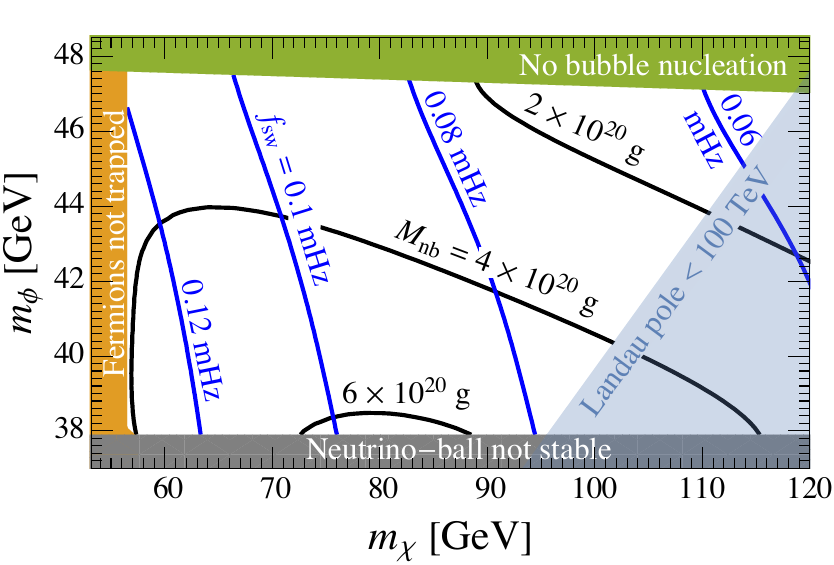}\figuretag{A1}
\caption{The parameter space fixing $\lambda_{\phi}=0.18$, $\lambda_\eta=0.30$, and $\lambda_{\phi\eta}=1.2$. The neutrino-ball mass and GW frequency are plotted as black and blue contours, respectively. The orange, gray, green, and light blue regions represent the parameter space that the mass gap is insufficient to trap $\chi$'s in the false vacuum, the solitons are unstable against $\chi\to\phi\nu$ decay, the bubbles cannot nucleate, and the Yukawa Landau pole is below 100 TeV, respectively. See the text for details.}\label{fig:n-ball}
\end{figure}

Below $T_\eta=(-\mu_\eta^2/c_\eta)^{1/2}$, the field configuration $(\phi,\eta)=(0,\pm u'_T)$ becomes the minimum of the thermal potential, where $u'_T=\sqrt{-(\mu_\eta^2+c_\eta T^2)/\lambda_\eta}$. In this case, the $\eta$-direction domain walls  can form with a surface tension of $\sigma_{\rm dw}=4\sqrt{\lambda_\eta/2}u'^3_T/3$. When the Universe cools below the critical temperature $T_c$, another minimum $(w_T,0)$ develops and becomes the global minimum, where $w_T=\sqrt{-(\mu_\phi^2+c_\phi T^2)/\lambda_\phi}$. At this stage, the Universe acquires a probability to decay into this true vacuum. Domain walls serve as the local impurities to seed the bubble nucleation, and the vacuum decay rate per area on the wall is
\be
\Gamma_S(T)\simeq\sigma_{\rm dw}e^{-S_{\rm inh}},\quad S_{\rm inh}=\frac{E_R}{T},
\ee
where $E_R$ is the height of the energy barrier of the bubble, evaluated by a spheroid bubble configuration under the thin-wall approximation~\cite{Blasi:2022woz}. Different from  the conventional homogeneous nucleation, the nucleation rate is now  given by the following integration
\be
\mN(T)=\int_T^{T_c}\frac{\d T'}{T'}\frac{\Gamma_S(T')}{H^3(T')},
\ee
and $\mN(T_n)=1$ sets the nucleation temperature $T_n$, which features $T_n\approx T_*$ under the thin-wall approximation.

Once $T_*$ is obtained, we can evaluate $m_ \chi^*/(\gamma_wT_{\rm rh})$ with $T_{\rm rh}\approx(1+\alpha)^{1/4}T_*$. If $m_ \chi^*/(\gamma_wT_{\rm rh})\gtrsim10$,  then the trapping fraction is close to 1 and neutrino-balls can form~\cite{Hong:2020est}, where   we obtain the neutrino-ball profiles and the relic abundance. Since $Y_ \chi=8Y_B/63$ is fixed by chemical equilibrium at $T_*<T<T_{\rm sph}$, we find that the effective mass of $\chi$ inside the neutrino-ball yields $\mu\approx37.9$ GeV  while  the neutrino-ball mass is around $10^{20}$ g. 

Following the guidelines of \Eq{analytical_Tc}, we can find the parameter space allowing for FOPT and neutrino-ball formation. An example is shown in Fig.~\ref{fig:n-ball}, where $\lambda_{\phi}=0.18$, $\lambda_\eta=0.30$, and $\lambda_{\phi\eta}=1.2$ are fixed, and we scan over the mass $m_\chi$ and $m_\phi$ in the true vacuum. The resultant neutrino-ball mass and GW frequency are shown in black and blue contours, respectively. Note that the numerical results deviate slightly from the general estimate in Fig.~\ref{fig:landscape}, as here we are considering a domain-wall-catalyzed FOPT that is different from the conventional homogeneous one. It is worthwhile to mention that initial estimates have suggested that Fermi-balls could collapse into PBHs during cooling via the increasing range of the interior Yukawa force~\cite{Kawana:2021tde,Marfatia:2021hcp,Huang:2022him,Marfatia:2022jiz,Lu:2022jnp,Tseng:2022jta,Kim:2023ixo,Borah:2024lml,Borah:2025wzl}, but more recent mean-field calculations indicate that this collapse is challenging for light Fermi-balls, because the effective scalar potential generated by the constituent fermions provides an intrinsic range of force that resists collapse~\cite{Xie:2024mxr}. Consequently, in our scenario, the relic from the FOPT is made up of  solitons instead of PBHs.

We see from  Fig.~\ref{fig:n-ball} that  if $m_\chi$ is too small, the mass gap is insufficient to trap $\chi$'s in the false vacuum, resulting in the orange region. If $m_\phi<\mu$, the constituent $\chi$ particles  inside the neutrino-ball can decay via $\chi\to\phi\nu$ rapidly, leading to unstable solitons. This is shown by the gray region. Conversely, if $m_\phi$ is too large, the bubble nucleation probability decreases so that  the FOPT is prevented, as indicated in the green region. Finally, although a sizable $y_\phi$ is favored for fermion trapping, a too large $y_\phi$ causes the breakdown of perturbativity. We use the light blue region to indicates values of $y_\phi\gtrsim1.5$, where the Landau pole of this Yukawa coupling is below 100 TeV.

\bibliographystyle{utphys}
\bibliography{references}

\end{document}